# Investigation into Pulsar Glitch Parameters

Miracle Chibuzor Marcel
Department of Physics and Astronomy, University of Nigeria, Nsukka, Enugu State, Nigeria
miracle.c.marcel@gmail.com

**Abstract**

An updated analysis of pulsar glitch parameters was conducted using a sample of 215 pulsars, encompassing 677 recorded glitch events, each glitch at least once between 1968 and 2024. The glitch rates were estimated and plotted against various pulsar parameters, including rotational frequency, spin-down rate, and characteristic age. These relationships were analyzed using Pearson's correlation coefficient. The results indicate linear and weak relationships between glitch rate and the pulsar parameters. However, the relationship between glitch rate and characteristic age is inversely weak.

## 1.0 Introduction

A neutron star is a celestial object composed of densely packed degenerate neutrons (Shapiro et al., 1983). These stars form at the core of supernova explosions in high-mass stars, where electrons and protons are compressed by the weight of infalling matter to form neutrons (Kutschera, 1998; Baade and Zwicky, 1934; Zwicky, 1958, 1974). The first neutron star was discovered by Jocelyn Bell Burnell in 1967 (Burnell & Bell, 1977). "Neutron star" is a general term encompassing various exotic celestial bodies, including pulsars (Shapiro et al., 1983), magnetars (Negreiros et al., 2018), anomalous X-ray pulsars (AXPs) (Van Paradijs, 1995), and millisecond pulsars (MSPs) (Bhattacharya, 1991).

Some important characteristics of neutron stars include the following: they are typically the size of a small city, with a radius of about 10-14 km and a mass between 1.4 and 3 solar masses (Potekhin, 2010). Their temperature ranges from hundreds of thousands to millions of degrees Kelvin (Lattimer, 2015). Neutron stars have extremely strong magnetic fields, ranging from $10^4$ to $10^{11}$ tesla (Reisenegger, 2003). Their surface gravity is around $2.0 \times 10^{12}$ m/s², approximately $2 \times 10^{11}$ times stronger than Earth's (Burnell, 2004).

Neutron stars can also experience "starquakes," known as glitches, during which the star's rotational frequency (F0) suddenly increases (a "spin-up") and then gradually decreases due to braking from the emission of radiation and high-energy particles (Wang et al., 2001; Espinoza et al., 2011; Yu et al., 2013; Watts et al., 2014). Pulsar glitches are believed to originate from internal processes within the neutron star's core. Studying these glitches is crucial to astrophysics, as they provide the only observable phenomena arising from the star's interior, offering insights into the state of matter within a neutron star's core (Eya & Urama, 2014).

Several studies have examined pulsar glitches. Espinoza et al. (2011) investigated the relationship between glitch rates and the characteristic ages of pulsars, finding that middle-aged pulsars (τc ~ 10 thousand years) experienced the most glitches, while older pulsars (τc ≥ 20 million years) and newborn pulsars (τc ≤ 1 thousand years) exhibited fewer glitches. They also observed a similar trend in the glitch spin-up rate (frequency derivative). Their study utilized pulsar data from the Jodrell Bank Observatory database, analyzing 315 glitches across 102 pulsars that glitched at least, including radio, X-ray, anomalous X-ray, and millisecond pulsars. Eya & Urama (2014) expanded on these studies using the same dataset, providing more precise measurements summarized in the table below.

Table 1. Glitch rate in different age bracket

| Age (years) | Number of Pulsars | Number of Glitches/ Glitch rate | Number of Glitch per Pulsar |
|---|---|---|---|
| $10^1 - 10^3$ | 1 | 2 | 2.0 |
| $10^3 - 10^5$ | 38 | 196 | 5.2 |
| $10^5 - 10^7$ | 59 | 112 | 1.9 |
| $10^7 - 10^9$ | 4 | 5 | 1.3 |
| Total | 102 | 315 | 10.4 |

Table retrieved from Eya & Urama (2014)

A more recent study by Millhouse et al. (2022), using all available pulsar data (including both pulsars that have glitched and those that have not) from the Jodrell Bank Observatory database (Espinoza et al. 2011; Basu et al. 2022) and the Australian Telescope National Facility Pulsar Catalog (Manchester et al. 2005), has shown that the trend observed by Espinoza et al. (2011) still holds.

This study aims to provide an updated analysis of the relationship between glitch rate and pulsar rotational parameters, including frequency, frequency derivative, spin-down rate, and characteristic age. It also serves as an update to a study conducted by the author in 2018, originally completed as part of the requirements for a Bachelor of Science (B.Sc.) degree in Physics and Astronomy at the University of Nigeria, Nsukka, and is now prepared for publication.

**2.0 Methodology**

On November 8, 2024, after conducting a literature review, the Australian Telescope National Facility Pulsar Catalog website (https://www.atnf.csiro.au/research/pulsar/psrcat/) was consulted and queried. Some selections were made including the pulsar name (based on J2000 coordinates), frequency F0 (Hz), frequency derivative F1 (s⁻²), discovery date, number of glitches per pulsar, and age of each pulsar. The total number of pulsars selected was 215, with 677 recorded glitch events, each having glitched at least once between 1968 and November 8, 2024, corresponding to 57789.5 in Modified Julian Date (MJD). The data were then imported into Google Sheets for further input and analysis as shown in Table 2 at the Data Availability section.

To compute the glitch rate for each pulsar, the formula Glitch Rate = $\frac{N}{T_{Obs}}$ was applied, where N is the number of glitches per pulsar and $T_{Obs}$ is the observational period. To determine $T_{Obs}$, following Millhouse et al. (2022), who defined $T_{Obs}$ as the time interval between the date a particular pulsar's discovery was reported in a publication (as listed in the ATNF catalog) and a chosen end date. For this study, the end date, November 8, 2024, corresponding to 57789.5 in MJD was used, which is the exact date of the retrieval of the data for this study. $T_{Obs}$ was calculated by subtracting the discovery year from 2024.8552 (decimal conversion), then computed the glitch rate accordingly.

Logarithmic transformations were performed on the glitch rate, frequency, frequency derivative, and characteristic age, and finally, the data were transferred to the OriginPro program to generate plots.

### 3.0 Results

Two plots were made for each relationship: one without the log and another with the log. The logarithm function compresses large values and expands small values. Consequently, data points that were previously close together on a linear scale in the first plot are spread out on a logarithmic scale in the second plot.

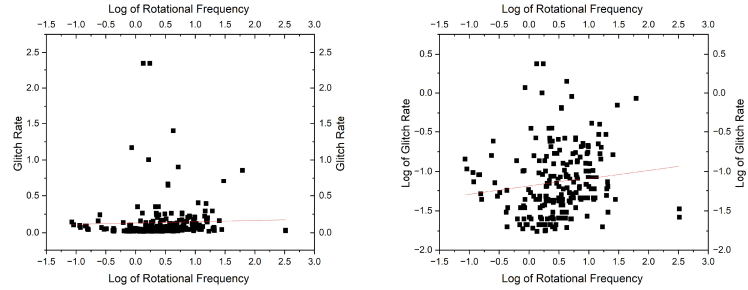

**Figures 1:** Plots of glitch rate vs. the logarithm of rotational frequencies for the 215 pulsars.

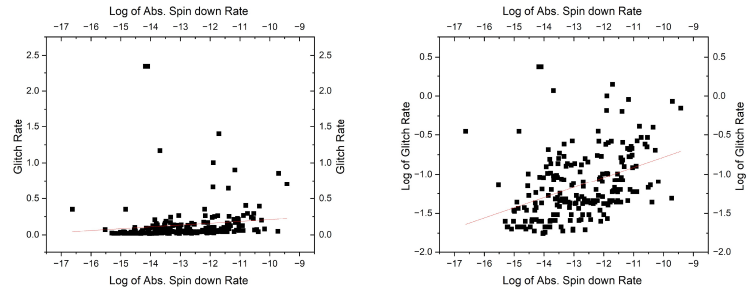

**Figures 2:** Plot showing the relationship between glitch rate and the logarithm of the absolute spin-down rate for the 215 pulsars.

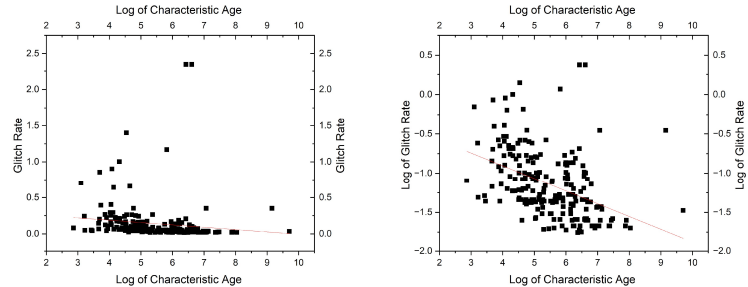

**Figure 3:** Plots showing the relationship between glitch rate and the logarithm of the characteristic ages for the 215 pulsars.

### 4.0 Discussions

Similar studies to the one presented here have been conducted in the past by Link et al. (1999), Urama & Okeke (1999), Lyne et al. (2000), Wang et al. (2000), Andersson et al. (2012), and Eya et al. (2019a, 2022). However, our study is unique because we incorporated all available data from 1968 to 2024, including data for newly discovered pulsars.

Pearson's correlation coefficient (r) (Cohen et al. 2009) measures the linear relationship between two variables, with a value ranging from -1 to 1. An r value of 1 indicates a perfect positive linear relationship, meaning that as one variable increases, the other also increases. An r value of -1 indicates a perfect negative linear relationship, where as one variable increases, the other decreases. An r value of 0 means no linear relationship. Therefore, the closer the value of r is to 1 or -1, the stronger the linear relationship.

Based on the above statement, we analyze our results. The correlation coefficient between the glitch rate and the log of rotational frequency is 0.03, while the correlation coefficient between the log of glitch rate and the log of rotational frequency is 0.13. These low correlation coefficients suggest very weak linear relationships. Specifically, the near-zero value of 0.03 indicates almost no linear correlation between glitch rate and the log of rotational frequency. Similarly, the value of 0.13 suggests a slight but weak positive linear relationship between the log of glitch rate and the log of rotational frequency. Consequently, changes in rotational frequency do not strongly predict changes in glitch rate.

The correlation coefficient between the glitch rate and the absolute spin-down rate is 0.12, while the correlation coefficient between the log of the glitch rate and the log of the absolute spin-down rate is 0.40. Both correlation coefficients indicate weak relationships. Specifically, the value of 0.12 suggests a very weak linear relationship between the glitch rate and the absolute spin-down rate. The higher value of 0.40, while still indicating a weak relationship, shows that the log-transformed data has a slightly stronger linear association. Overall, these results suggest that changes in the absolute spin-down rate do not strongly predict changes in the glitch rate.

The correlation coefficient between the glitch rate and the log of characteristic age is -0.13, while the correlation coefficient between the log of the glitch rate and the log of characteristic age is -0.42. These negative values suggest that the two quantities are inversely related to each other. Specifically, the value of -0.13 indicates a weak inverse linear relationship between the glitch rate and the log of characteristic age. The stronger negative value of -0.42 suggests a more pronounced inverse relationship when both variables are log-transformed. This means that as the characteristic age increases, the glitch rate tends to decrease, and this relationship is more evident when using the logarithmic scale.

### 5.0 Conclusion

With the updated data, new knowledge has been contributed by demonstrating that the relationships between glitch rate and pulsar rotational parameters, specifically rotational frequency, spin-down rate, and characteristic age are very weak on a linear scale. Additionally, the relationship between glitch rate and characteristic age remains inversely correlated. These findings are consistent with the works of several authors, including Link et al. (1999), Urama & Okeke (1999), Lyne et al. (2000), Wang et al. (2000), Andersson et al. (2012), and Eya et al. (2019a, 2022). We recommend exploring polynomials of higher ranks to better fit the data.

**Acknowledgment**

The author is grateful to the Jodrell Bank Centre for Astrophysics, the University of Manchester, and the Australia Telescope National Facility for the wealth of information on pulsar catalogs made available on their websites.

Special appreciation goes to the anonymous referee for their valuable comments, which greatly improved this paper.

The author is deeply indebted to the Department of Physics and Astronomy, University of Nigeria, for nurturing him in both character and learning and for being home to a legion of remarkable intellectuals who were all helpful in various ways throughout his academic journey at the university.

**DATA AVAILABILITY**

The data underlying this article are from the Australian Telescope National Facility Pulsar Catalogue (Manchester et al. 2005) and can be accessed at https://www.atnf.csiro.au/research/pulsar/psrcat/. The spreadsheet used as a table of values can be accessed here.